\documentclass[twocolumn,showpacs,preprintnumbers,amsmath,amssymb,superscriptaddress,nofootinbib,english]{revtex4-1}
\usepackage{times,amsmath,amsfonts,amssymb,epstopdf}
\usepackage{graphicx}
\usepackage{dcolumn}
\usepackage{bm}
\usepackage{epsfig}
\usepackage{graphicx}
\usepackage{hyperref}
\usepackage[usenames]{color}
\usepackage{url}
\usepackage[normalem]{ulem}
\usepackage[T1]{fontenc}
\usepackage[dvipsnames]{xcolor}

\newcommand{\rd}{{\rm d}}

\def\be{\begin{equation}}
\def\ee{\end{equation}}
\def\ba{\begin{eqnarray}}
\def\ea{\end{eqnarray}}
\begin{document}

\title{Exploring Vainshtein mechanism on adaptively refined meshes}

\author{Baojiu~Li}
\email[Email address: ]{baojiu.li@durham.ac.uk}
\affiliation{Institute for Computational Cosmology, Department of Physics, Durham University, Durham DH1 3LE, UK}

\author{Gong-Bo~Zhao}
\email[Email address: ]{gong-bo.zhao@port.ac.uk}
\affiliation{Institute of Cosmology and Gravitation, University of Portsmouth, Portsmouth PO1 3FX, UK}
\affiliation{National Astronomy Observatories, Chinese Academy of Science, Beijing, 100012, P.~R.~China}

\author{Kazuya~Koyama}
\email[Email address: ]{kazuya.koyama@port.ac.uk}
\affiliation{Institute of Cosmology and Gravitation, University of Portsmouth, Portsmouth PO1 3FX, UK}

\date{\today}

\begin{abstract}

There has been a lot of research interest in modified gravity theories which utilise the Vainshtein mechanism to recover standard general relativity in regions with high matter density, such as the Dvali-Gabadadze-Porrati and Galileon models. The strong nonlinearity in the field equations of these theories implies that accurate theoretical predictions could only be made using high-resolution cosmological simulations. Previously, such simulations were usually done on regular meshes, which limits both their performance and the accuracy. In this paper, we report the development of a new algorithm and code, based on {\sc ecosmog}, that uses adaptive mesh refinements to improve the efficiency and precision in simulating the models with Vainshtein mechanism. We have made various code tests against the numerical reliability, and found consistency with previous simulations. We also studied the velocity field in the self-accelerating branch of the DGP model. The code, parallelised using {\sc mpi}, is suitable for large cosmological simulations of Galileon-type modified gravity theories.

\end{abstract}


\maketitle

\section{Introduction}

\label{sect:introduction}

Modified gravity theories \cite{cfps2011} are proposed as alternative to the dark energy scenarios \cite{cst2006} to explain the observed accelerating expansion of our Universe \cite{gsetal2010,pretal2009,bbetal2011,rsetal2012,hletal2012,rmetal2009}, and have attracted a lot of research interest recently. Instead of invoking some sort of mysterious new energy component which drives the dynamics of the cosmos, these theories suggest that the Universe is filled with only normal plus dark matter (which is usually assumed to be cold), but the law of gravity can be different from that of standard general relativity (GR) on large scales, leading to the speed-up of the expansion rate.

Since the law of gravity is considered as universal, a modification to GR on large scales almost necessarily implies corresponding changes in the behaviour on small scales. Any such changes from GR, however, is already highly constrained by numerous local tests of gravity \cite{w2006}, and this has rendered many models incompatible with experimental bounds. Thus, any viable modified gravity theory should have some mechanism by which such modifications are suppressed and GR is restored in high-density regions such as the Solar system, where those experiments have been carried out and the resulted bounds apply. Such mechanisms are commonly referred to as `screening mechanisms' in the literature. Obviously, to make the model theoretically appealing, the screening mechanism needs to be an inherent (rather than add-on) property of it, which comes from the dynamics of the theory. The screening effect means that gravity behaves in different manners in different environments, and such environmental dependence often boils down to a high degree of nonlinearity in the relevant field equations, which makes the study of such theories rather challenging and (numerically) demanding.

In most modified gravity theories studied so far, the modification to GR is realised by a dynamical scalar-type (spin-0) field, which mediates the modified gravitational force. These theories could be roughly divided into two classes. In the first class of such theories, the screening is achieved by nonlinear couplings of the scalar field to matter and/or a nonlinear potential governing the self-interaction of the scalar field. If the coupling and potential are appropriately specified, the scalar field may acquire a very heavy mass in dense regions, making it hardly propagating or mediating any force between matter particles, or extremely weakly coupled to matter such that the resulted modification to GR very small. The chameleon models \cite{kw2004,ms2007}, with $f(R)$ gravity model \cite{sf2010} (see also \cite{lb2007,hs2007,bbds2008}) as a special example, belong to the former case, while the dilaton \cite{bbds2010} and symmetron \cite{hk2010} models belong to the latter case. In these models, $f(R)$ gravity (mainly the model of \cite{hs2007}) has been the most well-studied, and there have been many works which studied in detail its structure formation in the nonlinear regime, with the aid of $N$-body simulations \cite{o2008,olh2008,sloh2009,lz2009,zmlhf2010,lz2010,zlk2011,zlk2011b,lh2011,lzk2012,lkzl2012,lzlk2012,jblzk2012,lhkzjb2012}. During this process, we have developed an efficient $N$-body code, {\sc ecosmog} \cite{ecosmog}, based on the publicly available code {\sc ramses} \cite{ramses}, which is massively parallelised using {\sc mpi} and therefore makes large simulations for $f(R)$ gravity possible. Using the generic parameterisation for modified gravity theories of the first class \cite{bdl2011,bdlw2012}, {\sc ecosmog} has been recently extended to simulate general chameleon, dilaton and symmetron cosmologies \cite{bdlwz2012,bdlwz2013}.

The other class of modified gravity theories involving scalar degrees of freedom, with the Dvali-Gabadadze-Porrati (DGP) brane-world model \cite{dgp} as the most well-known example, realises the screening using nonlinear derivative self-couplings of a scalar field. Here, the nonlinearity makes the gravitational fields of individual particle interfere so that the deviation from GR per particle is weakened in a collection of particles compared with that for an isolated particle. This is known as the Vainshtein mechanism \cite{vainshtein} in the literature, which was originally introduced in the context of massive gravity to suppress the extra helicity modes of massive graviton to recover GR in the massless limit. The Vainshtein mechanism occurs not only in the recently proposed nonlinear massive gravity \cite{massivegravity1,massivegravity2,cp2012} and braneworld models but also in more general setups, such as the Galileon model \cite{nrt2009,dev2009, Galileon1, Galileon2, Galileon3, Galileon4}. The cosmological implications of the Galileon model have been studied by many groups (see, e.g.,\cite{ck2009,sk2009,gs2010,dt2010,ndt2010,ags2010,bbd2011,al2012,al2012b,blbp2012,nrcpagb2013,blbp2013}) in details. However, to date studies of large-scale structure formation in the nonlinear regime, even for the simplest case of DGP, have been very limited. The primary reason is that the field equations for these theories generally involve high products of second-order derivatives (see below), which make them difficult to solve numerically. There have been a few works on $N$-body simulations for DGP \cite{schmidt2009,schmidt2009b,cs2009,kw2009}, but these are so far limited to codes which, unlike {\sc ecosmog}, have no mesh refinements. Furthermore, for the code to solve the DGP equation one may have to smooth the underlying density field to reduce the noise, which certainly limits the force resolution of the simulations.

In this paper, we report the development and initial results of a new $N$-body code, {\sc ecosmog-v}, which is a sibling version of {\sc ecosmog} but for simulations of theories involving the Vainshtein screening. We have made various tests of the code to be confident about its accuracy and reliability. Using this code, we have studied the matter and velocity density fields in the self-accelerating branch of the DGP model. We find good agreement of our matter power spectrum with that of previous work \cite{schmidt2009,cs2009}, but the code enables us to go to even smaller scales and discover new interesting features. We also find that, like in the case of $f(R)$ gravity \cite{lhkzjb2012}, the velocity field is more strongly affected by the modification to gravity law, and this provides a potentially powerful cosmological test of the theory of gravity.

The structure of this paper is as follows. In \S~\ref{sect:model} we briefly describe the DGP model and the Vainshtein screening mechanism, which lays down the essential theoretical background. In \S~\ref{sect:equations} we present the relevant field equations in the form as they appear in our numerical simulation code, and introduce the algorithm used to solve these equations as well as its implementation. In \S~\ref{sect:code_tests} we will show the various checks which we have performed to ensure that our code works accurately in different situations and aspects. We then run a suite of 36 simulations for the DGP model and its variants, and use these to study their predictions about the matter and velocity fields, especially on small scales which have been un-probed by previous studies: the results are given in \S~\ref{sect:results}. Finally we summarise and outlook in \S~\ref{sect:summary}.

Throughout the paper we shall follow the metric convention $(+,-,-,-)$, and set $c=1$ except in the expressions where $c$ appears explicitly.

\section{The DGP model}

\label{sect:model}

\subsection{The model}
As an example of models that accommodate the Vainshtein mechanism, we consider the self-accelerating branch of the DGP braneworld model because this is the model where the Vainshtein mechanism has been studied in details (e.g.,~\cite{ls2003,lss2004,twobody}). This makes it possible to compare our new simulations with previous studies in the literature without the mesh refinements and demonstrate the power of the adaptive mesh technique to study the nonlinear clustering. However, we should note that this branch of the DGP model suffers from the ghost instability (e.g. \cite{ghost}) and it cannot be considered as a physical model. Moreover, this model is already strongly disfavoured by observations, and therefore should be considered as a toy model to test our code. We will report results on more physical models as follow-up works in the future.

The Friedman equation in the self-accelerating branch is given by
\begin{equation}
H^2 = \frac{H}{r_{\rm c}} + \frac{8 \pi G }{3}\rho,
\end{equation}
in which $r_{\rm c}$ is the cross-over length scale at which gravity becomes five-dimensional, $G$ is Newton's constant on the four-dimensional brane where matter fields live and $\rho$ matter density. To reproduce the observed cosmic acceleration one typically needs $r_{c}\sim H_0^{-1}$. The Friedman equation can be rewritten as
\begin{equation}
E(a) \equiv \frac{H(a)}{H_0} = \sqrt{\Omega_{{\rm rc}}} + \sqrt{\Omega_m a^{-3} + \Omega_{{\rm rc}}},
\end{equation}
where $a$ is the scale factor, $a=1$ today, and
\begin{equation}
\Omega_{{\rm rc}} \equiv \frac{1}{4 H_0^2 r_{\rm c}^2}, \quad \Omega_m \equiv  \frac{8 \pi G}{3 H_0^2} \rho_{m0}.
\end{equation}
The subscript $_0$ indicates that the quantity is evaluated today. 

Under the quasi-static approximation, the Poisson equation and the equation for the scalar field are given by \cite{bending}
\begin{equation}
\nabla^2 \varphi + \frac{r_{\rm c}^2}{3 \beta(a) a^2}
[(\nabla^2 \varphi)^2 -(\nabla_i \nabla_j \varphi)(\nabla^i \nabla^j \varphi)]
= \frac{8 \pi G a^2}{3 \beta(a)} \rho \delta,
\end{equation}
and
\begin{equation}
\nabla^2 \Psi = 4 \pi G a^2 \rho \delta + \frac{1}{2} \nabla^2 \varphi,
\end{equation}
where $\Psi$ is the gravitational potential, $\nabla$ is the spatial gradient operator and $\delta$ is matter density contrast. The function $\beta(a)$ is defined by\footnote{We shall drop the time-dependence for $\beta$ in the text hereafter for brevity.}
\begin{equation}\label{eq:beta}
\beta = 1- 2 H r_{\rm c} \left(1+ \frac{\dot{H}}{3 H^2} \right),
\end{equation}
which can be written as
\begin{eqnarray}
\beta &\equiv& -\frac{\left(\frac{1}{2}\Omega_{\rm
M}a^{-3}+\Omega_{{\rm rc}}\right)}{\sqrt{\Omega_{{\rm rc}}\left(\Omega_{\rm M}a^{-3}+\Omega_{{\rm rc}}\right)}}.
\end{eqnarray}

If we linearise the equations, the Poission equation is given by
\begin{equation}
\nabla^2 \Psi = 4 \pi G a^2 \left(1+ \frac{1}{3 \beta} \right) \rho \delta.
\label{linear}
\end{equation}
Note that $\beta$ is always negative so the growth of structure formation is suppressed in this model.


\subsection{Vainshtein mechanism}

The relevant field equation in a spherically symmetric configuration can be written in the following form
\begin{eqnarray}\label{eq:dgp}
&&\frac{2r_{\rm c}^2}{3\beta a^2}\frac{1}{r^2}\frac{\rd}{\rd r}\left[r\left(\frac{\rd\varphi}{\rd r}\right)^2\right] + \frac{1}{r^2}\frac{\rd}{\rd r}\left[r^2\frac{\rd\varphi}{\rd r}\right]\nonumber\\
&=& \frac{8\pi G}{3\beta}\delta\rho_ma^2,
\end{eqnarray}
in which $\delta\rho_m=\rho\delta$ is the matter density perturbation and $\varphi$ is the scalar field (i.e., the brane bending mode), and $\beta$ is given in Eq.~(\ref{eq:beta}). 
Defining the mass enclosed in radius $r$ as
\begin{eqnarray}
M(r) &\equiv& 4\pi\int^r_0\delta\rho_m(r')r'^2dr',
\end{eqnarray}
we can rewrite Eq.~(\ref{eq:dgp}) as
\begin{eqnarray}\label{eq:dgp_eqn}
\frac{2r_{\rm c}^2}{3\beta}\frac{1}{r}\left(\frac{\rd\varphi}{\rd r}\right)^2 + \frac{\rd\varphi}{\rd r} &=& \frac{2}{3\beta}\frac{GM(r)}{r^2}\ \equiv\ \frac{2}{3\beta}g_N(r),
\end{eqnarray}
where for simplicity we have set $a=1$, and $g_N$ is the Newtonian acceleration caused by the mass $M(r)$ at distance $r$ from the centre.

To further simplify the situation, let us assume that $\delta\rho_m$ is constant within radius $R$ and zero outside. Then Eq.~(\ref{eq:dgp_eqn}) has the physical solution
\begin{eqnarray}\label{eq:dvarphidr_in}
\frac{\rd\varphi}{\rd r} &=& \frac{4}{3\beta}\frac{r^3}{r_\ast^3}\left[\sqrt{1+\frac{r_\ast^3}{r^3}}-1\right]g_N(r)
\end{eqnarray}
for $r\geq R$ and
\begin{eqnarray}\label{eq:dvarphidr_out}
\frac{\rd\varphi}{\rd r} &=& \frac{4}{3\beta}\frac{R^3}{r_\ast^3}\left[\sqrt{1+\frac{r_\ast^3}{R^3}}-1\right]g_N(r)
\end{eqnarray}
 for $r\leq R$. Here we have defined the Vainshtein radius $r_\ast$ by
 \begin{eqnarray}
 r_\ast^3 \equiv \frac{8r^2_cr_s}{9\beta^2},
 \end{eqnarray}
where $r_s\equiv2GM$ is the Schwarzschild radius.

The fifth force is given by $\frac{1}{2}\rd\varphi/\rd r$. So when $r\gg r_\ast$ we have
\begin{eqnarray}
\frac{1}{2}\frac{\rd\varphi}{\rd r} &\rightarrow& \frac{1}{3\beta}g_N(r),
\end{eqnarray}
meaning that above the Vainshtein radius gravity is weakened (because $\beta<0$ for the self-accelerating branch of the DGP model). On the other hand, when $r\ll r_\ast$ we have
\begin{eqnarray}
\frac{1}{2}\frac{\rd\varphi}{\rd r} &\rightarrow& 0,
\end{eqnarray}
indicating that the fifth force gets suppressed well within the Vainshtein radius.

As some useful examples, we have listed in the Table~\ref{table:simulations} the different radii of some objects. We could see that from objects as small as atoms to those as large as the Milky Way, the Vainshtein radius is significantly larger than the physical size, and as a result the fifth force is negligible for those objects.
\begin{table}
\caption{The radii $R$ (physical size), $r_{\rm s}$ (Schwarzschild radius) and $r_\ast$ (Vainshtein radius) of some typical objects. Unit is meter. For estimation we have used $\beta=2\sqrt{2}/3$ and $r_{\rm c}=H_0^{-1}\sim4,000{\rm Mpc}\sim1.2\times10^{26}$m.}
\begin{tabular}{@{}lccc}
\hline\hline
Object & $R$ & $r_{\rm s}$ & $r_\ast$\\
\hline
Universe & $\sim1.2\times10^{26}$ & $\sim4.5\times10^{25}$ & $\sim8.6\times10^{25}$\\
Milky Way & $\sim0.9\times10^{21}$ & $\sim2\times10^{15}$ & $\sim3\times10^{22}$\\
Sun & $\sim0.7\times10^{9}$ & $\sim3\times10^3$ & $\sim3.5\times10^{18}$\\
Earth & $\sim6\times10^{6}$ & $\sim9\times10^{-3}$ & $\sim5\times10^{16}$\\
Atom & $\sim5\times10^{-11}$ & $\sim1.8\times10^{-54}$ & $\sim3\times10^{-1}$\\
\hline
\end{tabular}
\label{table:simulations}
\end{table}

\section{$N$-body equations and algorithm}

\label{sect:equations}

In this section we describe the $N$-body equations in appropriate units and their discretised versions which the {\sc ecosmog} code solves to do cosmological simulations.

\subsection{$N$-body equations}

The code unit used in our code is based on the supercomoving coordinates of \cite{ms1998,ramses}, which can be summarised as follows (tilded quantities are expressed in the code unit):
\begin{eqnarray}\label{eq:code_unit}
\tilde{x}\ =\ \frac{x}{aB},\ \ \ \tilde{\rho}\ =\ \frac{\rho a^3}{\rho_c\Omega_m},\ \ \ \tilde{v}\ =\ \frac{av}{BH_0},\nonumber\\
\tilde{\phi}\ =\ \frac{a^2\phi}{(BH_0)^2},\ \ \ d\tilde{t}\ =\ H_0\frac{dt}{a^2},\ \ \ \tilde{\varphi}\ =\ \frac{c^2a^2\varphi}{(BH_0)^2},
\end{eqnarray}
in which $x$ is the physical coordinate, $\rho_c$ is the critical density today, $\Omega_m$ the fractional energy density for matter today, $v$ the particle velocity, $\phi$ the gravitational potential and $c$ the speed of light. In addition, $B$ is the size of the simulation box in unit of $h^{-1}$Mpc and $H_0$ the Hubble expansion rate today in unit of $100h$~km/s/Mpc. Note that with these conventions the average matter density is $\tilde{\bar{\rho}}=1$. All these quantities are dimensionless.

Using these quantities, the EOM of the brane-bending mode can be written as
\begin{eqnarray}\label{eq:code_eqn}
\tilde{\nabla}^2\tilde{\varphi} + \frac{R^2_{\rm c}}{3\beta a^4}\left[\left(\tilde{\nabla}^2\tilde{\varphi}\right)^2-\left(\tilde{\nabla}_i\tilde{\nabla}_j\tilde{\varphi}\right)^2\right] &=& \frac{\Omega_m a}{\beta}\left(\tilde{\rho}-1\right)\ \ \ \ \
\end{eqnarray}
in which we have defined a new dimensionless $\mathcal{O}(1)$ quantity
\begin{eqnarray}
R_{\rm c} &\equiv& \frac{H_0r_{\rm c}}{c}\ =\ \frac{1}{2\sqrt{\Omega_{\rm rc}}}.
\end{eqnarray}
For simplicity, from here on we shall ignore the tildes on all quantities in Eq.~(\ref{eq:code_unit}), and unless otherwise stated these quantities are all in code unit where they appear.

Eq.~(\ref{eq:code_eqn}) can be regarded as a second-order algebraic equation for $\nabla^2\varphi$, which has two branches of solutions. To avoid undecidedness in which branch of solution (which may cause numerical problems in the code) to take, we first solve it analytically and get
\begin{eqnarray}\label{eq:two_branch}
\nabla^2\varphi &=& \frac{1}{2(1-w)}\left[-\alpha\pm\sqrt{\alpha^2+4(1-w)\Sigma}\right],
\end{eqnarray}
where we have defined
\begin{eqnarray}
\alpha &\equiv& \frac{3\beta a^4}{R^2_{\rm c}},\\
\Sigma &\equiv& \left(\nabla_i\nabla_j\varphi\right)^2 - w\left(\nabla^2\varphi\right)^2 + \frac{\alpha}{\beta}\Omega_ma\left(\rho-1\right),
\end{eqnarray}
in which we have followed \cite{cs2009} and used the operator-splitting trick to avoid the numerical problem due to imaginary square root. $w$ is a constant coefficient which is chosen to be $1/3$.

Obviously, to get meaningful result in the limit $\rho\rightarrow1$, one has to take the ${\rm sign}(\alpha)$-branch of Eq.~(\ref{eq:two_branch}), where ${\rm sign}(\alpha)$ is the signature of $\alpha$ or equivalently of $\beta$, namely
\begin{eqnarray}
\label{eq:one_branch}
\nabla^2\varphi &=& \frac{1}{2(1-w)}\left[-\alpha+\frac{\alpha}{|\alpha|}\sqrt{\alpha^2+4(1-w)\Sigma}\right].
\end{eqnarray}
If a wrong branch is taken, then $\nabla^2\varphi\neq0$ even when $\rho\rightarrow1$, i.e., when the matter density field becomes uniform, and this is clearly a sign of inconsistency (indeed one of our code tests below is related to this observation).

The equation for the standard Newtonian potential, on the other hand, is the same as in GR, and we have \cite{ramses,ecosmog}
\begin{eqnarray}\label{eq:newton}
\nabla^2\phi &=& \frac{3}{2}\Omega_ma(\rho-1).
\end{eqnarray}
Once both $\phi$ and $\varphi$ are solved, the total gravitational potential in the DGP model, $\Psi= \phi+\varphi/2$, follows, which can be differenced to find the total force under which particles move.

\subsection{Discretisation}

Of course, computers could only handle a finite number of operations, and therefore Eqs.~(\ref{eq:one_branch}, \ref{eq:newton}) must be discretised before being put into a code to be solved. This is the task for this subsection.

Consider that we want to solve these equations on a three-dimensional mesh consisting of periodic cubic cells, with cell length $h$, and denote the value of the scalar field at the centre of the $(i,j,k)$-th cell by $\varphi_{i,j,k}$ (and similarly for other quantities), then the discrete version of the field derivatives are
\begin{eqnarray}
\nabla\varphi &=& \frac{1}{2h}\left(\varphi_{i+1,j,k}-\varphi_{i-1,j,k}\right),\\
\nabla^2\varphi &=& \frac{1}{h^2}\left(\varphi_{i+1,j,k}+\varphi_{i-1,j,k}-2\varphi_{i,j,k}\right),\\
\nabla_x\nabla_y\varphi &=& \frac{1}{4h^2}\Big(\varphi_{i+1,j+1,k}+\varphi_{i-1,j-1,k}-\varphi_{i+1,j-1,k}\nonumber\\
&& ~~~~~~~~~~-\varphi_{i-1,j+1,k}\Big),
\end{eqnarray}
where we have assumed one dimension for simplicity for $\nabla\varphi$ and $\nabla^2\varphi$, the three-dimensional generalisations of which are trivial.

Using these, it is straightforward to write down the discrete version of the Poisson equation as
\begin{widetext}
\begin{eqnarray}
\frac{1}{h^2}\Big(\phi_{i+1,j,k}+\phi_{i-1,j,k}+\phi_{i,j+1,k}+\phi_{i,j-1,k}+\phi_{i,j,k+1}+\phi_{i,j,k-1}-6\phi_{i,j,k}\Big) &=& \frac{3}{2}\Omega_m(\rho_{i,j,k}-1)\ \equiv\ \frac{3}{2}\Omega_m\delta_{i,j,k}.
\end{eqnarray}
\end{widetext}
Similarly, the EOM for the brane-bending mode can be written as an operator equation $\mathcal{L}^h\varphi_{i,j,k}=0$, with
\begin{widetext}
\begin{eqnarray}
\mathcal{L}^h\varphi_{i,j,k} &\equiv& \frac{1}{h^2}\Big(\varphi_{i+1,j,k}+\varphi_{i-1,j,k}+\varphi_{i,j+1,k}+\varphi_{i,j-1,k}+\varphi_{i,j,k+1}+\varphi_{i,j,k-1}-6\varphi_{i,j,k}\Big)\nonumber\\
&& - \frac{1}{2(1-w)}\left[-\alpha+\frac{\alpha}{|\alpha|}\sqrt{\alpha^2+4(1-w)\Sigma_{i,j,k}}\right],\ \ \ \
\end{eqnarray}
\end{widetext}
where the superscript $^h$ is used to label the level of the mesh (or equivalently the size of the cell of that level), and we have defined
\begin{widetext}
\begin{eqnarray}
\Sigma_{i,j,k} &\equiv&\frac{1-w}{h^4}\bigg[\Big(\varphi_{i+1,j,k}+\varphi_{i-1,j,k}-2\varphi_{i,j,k}\Big)^2+\Big(\varphi_{i,j+1,k}+\varphi_{i,j-1,k}-2\varphi_{i,j,k}\Big)^2+\Big(\varphi_{i,j,k+1}+\varphi_{i,j,k-1}-2\varphi_{i,j,k}\Big)^2\bigg]\nonumber\\
&& - \frac{2}{h^4}w\left(\varphi_{i+1,j,k}+\varphi_{i-1,j,k}-2\varphi_{i,j,k}\right)\left(\varphi_{i,j+1,k}+\varphi_{i,j-1,k}-2\varphi_{i,j,k}\right)\nonumber\\
&& - \frac{2}{h^4}w\left(\varphi_{i+1,j,k}+\varphi_{i-1,j,k}-2\varphi_{i,j,k}\right)\left(\varphi_{i,j,k+1}+\varphi_{i,j,k-1}-2\varphi_{i,j,k}\right)\nonumber\\
&& - \frac{2}{h^4}w\left(\varphi_{i,j+1,k}+\varphi_{i,j-1,k}-2\varphi_{i,j,k}\right)\left(\varphi_{i,j,k+1}+\varphi_{i,j,k-1}-2\varphi_{i,j,k}\right)\nonumber\\
&& + \frac{1}{8h^4}\Big(\varphi_{i+1,j+1,k}+\varphi_{i-1,j-1,k}-\varphi_{i+1,j-1,k}-\varphi_{i-1,j+1,k}\Big)^2\nonumber\\
&& + \frac{1}{8h^4}\Big(\varphi_{i+1,j,k+1}+\varphi_{i-1,j,k-1}-\varphi_{i+1,j,k-1}-\varphi_{i-1,j,k+1}\Big)^2\nonumber\\
&& + \frac{1}{8h^4}\Big(\varphi_{i,j+1,k+1}+\varphi_{i,j-1,k-1}-\varphi_{i,j+1,k-1}-\varphi_{i,j-1,k+1}\Big)^2 + \frac{\alpha}{\beta}\Omega_m\delta_{i,j,k}.
\end{eqnarray}
\end{widetext}
This equation can be solved using the Newton-Gauss-Seidel relaxation method, for which the code iterates to update the value of $\varphi_{i,j,k}$ in all cells, and at each iteration the field values changes as
\begin{eqnarray}
\varphi^{h,{\rm new}}_{i,j,k} &=& \varphi^{h,{\rm old}}_{i,j,k} - \frac{\mathcal{L}^h\left(\varphi^{h,{\rm old}}_{i,j,k}\right)}{\frac{\partial\mathcal{L}^h\left(\varphi^{h,{\rm old}}_{i,j,k}\right)}{\partial\varphi_{i,j,k}^{h,{\rm old}}}},
\end{eqnarray}
where we have
\begin{widetext}
\begin{eqnarray}
&&\frac{\partial\mathcal{L}^h\left(\varphi^{h,{\rm old}}_{i,j,k}\right)}{\partial\varphi_{i,j,k}^{h,{\rm old}}}\nonumber\\
&=& -\frac{6}{h^2} + \frac{\alpha}{|\alpha|} \frac{4(1-3w)}{h^4\sqrt{\alpha^2+4(1-w)\Sigma_{i,j,k}}}\Big(\varphi_{i+1,j,k}+\varphi_{i-1,j,k}+\varphi_{i,j+1,k}+\varphi_{i,j-1,k}+\varphi_{i,j,k+1}+\varphi_{i,j,k-1}-6\varphi_{i,j,k}\Big).\ \ \
\end{eqnarray}
\end{widetext}

Note that, in comparison with the corresponding equations in $f(R)$ gravity \cite{ecosmog}, here the nonlinearity mainly exists in the term $\Sigma_{i,j,k}$, which involves products of the derivatives. Such a special property of the Vainshtein screening makes it more difficult for the Newton-Gauss-Seidel relaxations to converge.

\subsection{Numerical implementation}

As mentioned above, {\sc ramses} and therefore {\sc ecosmog} are AMR codes. The code starts from a regular mesh which covers the whole simulation domain, which we call the domain grid, and adaptively refines the mesh if the effective number of dark matter particles in a given cell exceeds certain threshold $N_{\rm pc}$. In this way, it gives sufficient resolution in high matter density regions and avoids wasting too much time on low density regions where resolution requirement is mild. The process of self-refinement goes on until finally all cells in the finest level do not satisfy the refinement criterion. Whenever a refinement is created, it is used to compute the force experienced by the particles which fall in its domain, and the boundary conditions can be set by using the values interpolated from coarser levels (see \cite{ecosmog} for more details).

The essential features of the code that are relevant for modified gravity simulations have been described in detail in \cite{ecosmog}. Here we only present the most relevant part for our Vainshtein simulations.

It is well known that, when using relaxation method to solve boundary value problems, the first few relaxations see the solution approach the true value quickly. The convergence then becomes slower as one gets closer to the true solution, making the method less efficient. This is because relaxation on the fine grid can only damp the short-wavelength Fourier components of the error, leaving behind the smooth long-wavelength components which cause a global poor convergence. The common method which is used to tackle this problem is the mutligrid technique \cite{multigrid}. The idea is simple: when the convergence rate becomes low on the fine grid, one {\it coarsifies} the discrete equation by interpolation, moves it to the coarser level and solves it there. The long-wavelength components of the error then decays quickly on the coarse grids, giving more accurate solution there, which can then be interpolated back to the fine grid. We follow the standard V-cycle in the arrangement of multigrid: the field equation is always solved from the finest to the coarsest grid, and then back. If convergence is not achieved (see below), further V-cycles are used.

The multigrid method can be easily applied to computations based on a regular mesh, since the corresponding coarser grids are also regular and satisfy periodic boundary conditions. For refinements, which generally have irregular shapes, however, much effort needs to be devoted to designing the corresponding coarser meshes and setting appropriate boundary conditions for them. One of the important features of {\sc ramses} and {\sc ecosmog} is that it solves the discrete equation using multigrid technique on {\it both} the domain grid {\it and} the refinements, which significantly speeds up the convergence.

Consider the EOM for the brane-bending mode, which for simplicity can be written as
\begin{eqnarray}\label{eq:fine_grid_eqn}
\mathcal{L}^h\left(\varphi^h\right) &=& f^h
\end{eqnarray}
on the fine level, where $\mathcal{L}$ is the nonlinear operator acting on $\varphi^h$ defined above. Note that although we have $f^h=0$ for the DGP equation, we shall still keep $f^h$ in Eq.~(\ref{eq:fine_grid_eqn}) for the reason that will become clear soon.

After a certain number of pre-smoothing Gauss-Seidel iterations on the fine grid, the convergence becomes slow because short-wavelength modes of the error have already decayed and long-wavelength modes are untouched. One arrives at an approximate solution $\hat{\varphi}^h$ on the fine level, which gives
\begin{eqnarray}
\mathcal{L}^h\left(\hat{\varphi}^h\right)\ =\ \hat{f}^h \neq f^h.
\end{eqnarray}
Now consider what we call the residual equation,
\begin{eqnarray}
\mathcal{L}^h\left(\varphi^h\right) - \mathcal{L}^h\left(\hat{\varphi}^h\right)\ =\ f^h-\hat{f}^h\ \equiv\ -d^h,
\end{eqnarray}
where $d^h$ is the residual. After coarsifying and certain rearrangements, we obtain the coarser-level equation as
\begin{eqnarray}
\mathcal{L}^H\left(\varphi^H\right) &=& \mathcal{L}^H\left(\mathcal{R}\hat{\varphi}^h\right)-\mathcal{R}d^h,
\end{eqnarray}
where $\mathcal{R}$ is the {\it restriction} operator which compute the coarse-level values of quantities given their fine-level values by interpolation. After a number of relaxation iterations on the coarser level, we find an approximate solution $\hat{\varphi}^H$, which has long-wavelength modes of the error further reduced, and therefore the fine-level solution can be corrected as
\begin{eqnarray}
\hat{\varphi}^{h,{\rm new}} &=& \hat{\varphi}^{h,{\rm old}} + \mathcal{P}\left(\hat{\varphi}^H-\mathcal{R}\hat{\varphi}^h\right),
\end{eqnarray}
where $\mathcal{P}$ is the {\it prolongation} operator which does the opposite of the restriction operator, again by interpolation.

The {\it truncation error} $\tau^h$ can be estimated as
\begin{eqnarray}
\tau^h &\approx& \mathcal{L}^H\left(\mathcal{R}\hat{\varphi}^h\right) - \mathcal{RL}^h\left(\hat{\varphi}^h\right),
\end{eqnarray}
and similarly for other levels. This could provide a stopping (convergence) criterion for the multigrid iterations, which in our case is
\begin{eqnarray}
\left|d^h\right| \lesssim \alpha\left|\tau^h\right|,
\end{eqnarray}
in which $\alpha\leq1/3$ is some predefined constant, which is often set even smaller to be more conservative. Useful as it is, we find that this criterion is often not needed because it is easy to get $|d^{h}|$ down to $\mathcal{O}\left(10^{-10}\right)$ within a few to a few tens of iterations, particularly on the refinements. This is typically much smaller than the truncation error.

\section{Code tests}

\begin{figure*}
\includegraphics[scale=0.45]{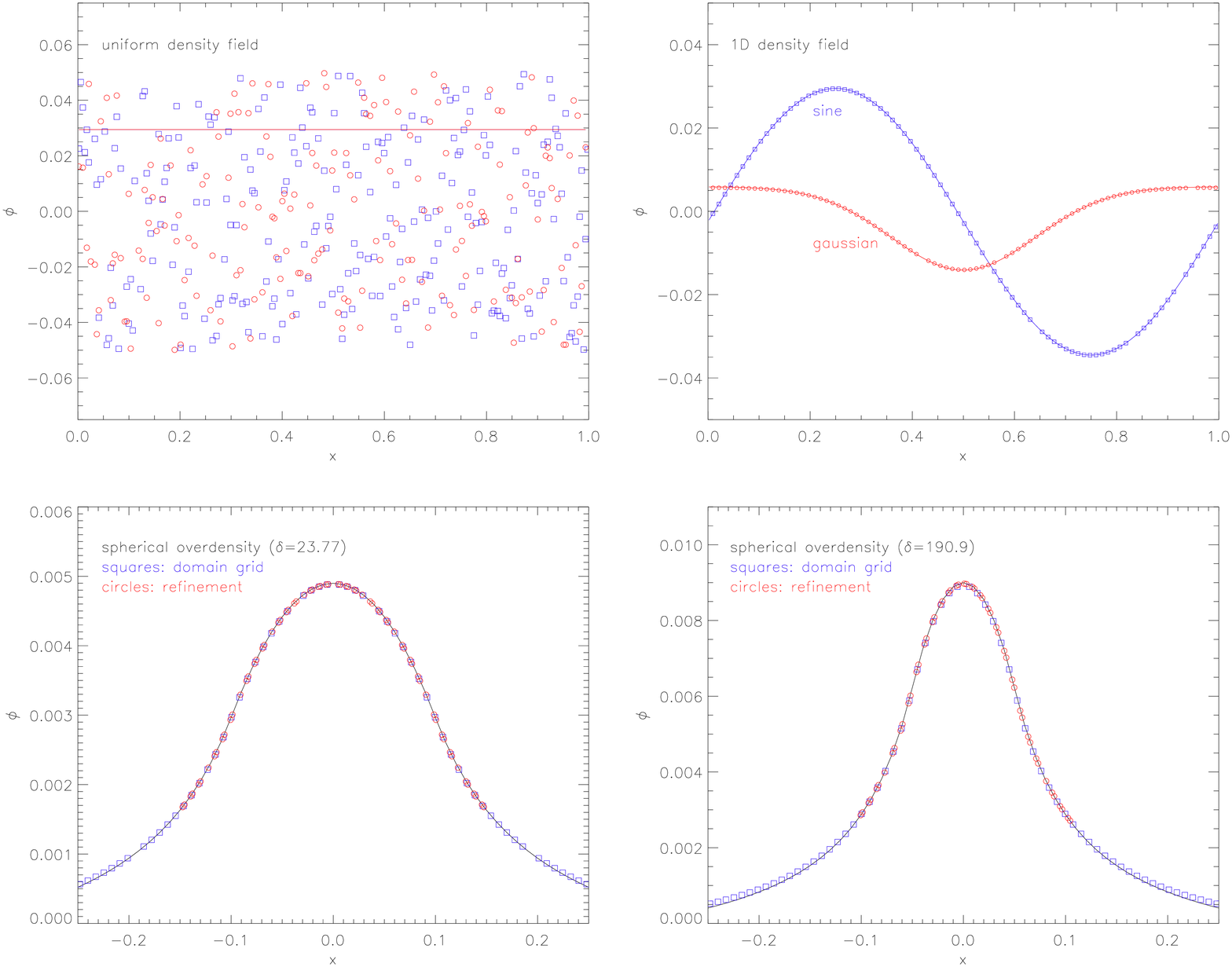}
\caption{(Colour online) Code tests. {\it Upper left panel}: test using a uniform density field and random initial guess for $\varphi(\mathbf{x})$ (\S~\ref{subsect:uniform_dens}); the symbols are the initial guesses and the solid lines are the final results of $\varphi$ after relaxation. {\it Upper right panel}: tests using 1D sine (blue squares) and Gaussian (red circles) density fields as described in Eqs.~(\ref{eq:delta_sine}, \ref{eq:delta_gaussian}) respectively (\S~\ref{subsect:1D_dens}); the solid curves are the analytical solutions. {\it Lower panels}: tests using a spherical overdensity with $\delta=23.77$ (left) and $\delta=190.9$ (right); the blue squares, red circles and black solid curves are respectively the code predictions on the domain grid, refinements and exact solutions obtained from numerical integration (\S~\ref{subsect:spherical_dens} and \S~\ref{subsect:multilevel}). In all tests we have assumed $\beta=-5.0, a=1.0$; the simulation box size is $128h^{-1}$Mpc and the domain grid has 256 cells in each dimension.} \label{fig:tests}
\end{figure*}

In this section we present the results of several tests of the {\sc ecosmog-v} code, which are essential for us to be confident about its reliability. For this purpose, we have performed several tests as shown in Fig.~\ref{fig:tests} (see the figure caption for more technical details).

\label{sect:code_tests}

\subsection{Uniform density field}

\label{subsect:uniform_dens}

The simplest test one can possibly do for the code is by considering a homogeneous distribution of matter, in which case the scalar field $\varphi$ remains a constant across the space, and its value can be arbitrary due to the shift symmetry in the EOM. If the code works correctly, then any inhomogeneity in $\varphi(\mathbf{x})$ should quickly disappear after a few Gauss-Seidel iterations.

To check this, we set $\rho_{i,j,k}=1$ and use random values that follow a uniform distribution between $-0.05$ and $0.05$ as the initial guess of $\varphi_{i,j,k}$. These are shown in the upper left panel of Fig.~\ref{fig:tests} as the symbols, in which for clarity we have fixed the $y$ and $z$ coordinates. We then let the code do the Gauss-Seidel relaxation until the residual gets small enough, $|d^h|<10^{-10}$. After a few iterations this criterion is satisfied and the resulted $\varphi(x)$ are described by the solid lines. We can see clearly that the final solution is constant in space.

We have shown two different initial guesses as blue squares and red circles respectively, and $\varphi(x)$ quickly converges to  a constant, which is almost the same in both cases.

\subsection{One dimensional density field}

\label{subsect:1D_dens}

In one dimension we can check that $(\nabla^2\varphi)^2=(\nabla_{i}\nabla_j\varphi)^2$, which means that the nonlinear term in the DGP equation simply vanishes. This provides another simple but very useful test of the code, because if there is something wrong in the nonlinear term in the code, then it is unlikely to vanish coincidentally and the solution for $\varphi$ should be incorrect.

Following \cite{ecosmog}, we do two tests using one-dimensional density field. The first test uses a sine-type field, specified by
\begin{eqnarray}\label{eq:delta_sine}
\delta(x) &=& -\frac{\beta}{\Omega_ma}K\sin(2\pi x),
\end{eqnarray}
where $K$ is a numerical coefficient which we take to be
\begin{eqnarray}
K &=& 0.128\pi^2.
\end{eqnarray}
Because the nonlinear term in the EOM does not contribute in 1D configurations, it can be easily checked that the analytical solution for this density field is
\begin{eqnarray}
\varphi(x) &=& 0.032 \sin(2\pi x).
\end{eqnarray}
In the upper right panel of Fig.~\ref{fig:tests}, we have presented this analytical solution (blue solid curve) together with the numerical solution from the code (blue squares), which agree with each other very well. We have tried a few other numerical values of $K$ and found same agreements.

The second test uses a Gaussian-type density field, given by
\begin{eqnarray}\label{eq:delta_gaussian}
\delta(x) &=& \frac{2J\alpha}{w^2}\left[1-\frac{(x-0.5)^2}{w^2}\right]\nonumber\\
&&~~~~~~~~~\times\exp\left[-\frac{(x-0.5)^2}{w^2}\right],
\end{eqnarray}
which corresponds to an exact analytic solution
\begin{eqnarray}
\varphi(x) &=& J\left[1-\alpha\exp\left(-\frac{(x-0.5)^2}{w^2}\right)\right].
\end{eqnarray}
Here $J, \alpha, w$ are constants which we take to be
\begin{eqnarray}
J\ =\ 0.02,~~\alpha\ =\ 0.9999,~~w\ =\ 0.2,
\end{eqnarray}
as a numerical example. In the upper right panel of Fig.~\ref{fig:tests} we show this analytical solution (red solid curve) together with the numerical solution given by our code (red circles): again there is a very good match.

Note that the code uses periodic boundary condition, which is exactly satisfied in the sine case. In the Gaussian case, the density field is not perfectly periodic, but $\delta(x)$ is sufficiently suppressed by the exponential factor in Eq.~(\ref{eq:delta_gaussian}), and is close to zero at $x=0$ and $x=1$, so that the periodic boundary condition approximately holds.

\subsection{Spherical overdensity}

\label{subsect:spherical_dens}

In the above we have tested the code for 1D density fields only. These tests show that the nonlinear term in the EOM do not create any problem, but they are not sufficient to prove that the code can solve this term correctly when it does contribute. For the latter we need to do tests in three dimensions, of which the spherically symmetric cases are the simplest possibility.

Following \cite{schmidt2009}, here we assume that the matter density inside the sphere is constant, i.e., we have a spherical overdensity. In code unit, Eqs.~(\ref{eq:dvarphidr_in}, \ref{eq:dvarphidr_out}) can be written as
\begin{eqnarray}
\frac{{\rm d}\varphi}{{\rm d}r} &=& -\frac{3\beta}{4R_{\rm c}^2}\left[\sqrt{1+\frac{8\Omega_mR^2_{\rm c}\delta }{9\beta^2}}-1\right]r,
\end{eqnarray}
for $r<R$ and
\begin{eqnarray}
\frac{{\rm d}\varphi}{{\rm d}r} &=& -\frac{3\beta}{4R_{\rm c}^2}\left[\sqrt{1+\frac{8\Omega_mR^2_{\rm c}\delta }{9\beta^2}\frac{R^3}{r^3}}-1\right]r,
\end{eqnarray}
for $r\geq R$, where $r$ is the comoving coordinate scaled by the boxsize $B$, while $R$ is the radius of the spherical overdensity also scaled by $B$. $\delta$ is the overdensity.

Given the value $\varphi(r=0)$, these equations can be integrated to find $\varphi(r>0)$ numerically. We do this for two different cases, respectively $\delta=23.77, R=0.05$ and $\delta=190.9, R=0.1$, which correspond to the same mass in the sphere, and the results are shown as the black solid curves in the lower panels of Fig.~\ref{fig:tests}. For comparison, solutions from our numerical code are shown as blue squares in those panels.

We can see that in both cases the two solutions agree very well, especially on small $r$. Far from the centre, the agreement becomes less perfect since the numerical integration does not assume periodicity of the spherical overdensity, while the numerical code uses period boundary condition so that the spherical density sees its own images. Such an artificial discrepancy due to imposing periodic boundary conditions can also be seen in the point-mass test of \cite{ecosmog}, and it emphasises the importance of choosing large enough box sizes in cosmological simulations.

\subsection{Multilevel}

\label{subsect:multilevel}

One of the most important features of our code compared with existing codes in the literature is that it enables adaptive mesh refinements to improve on the resolution. It is therefore crucial to check that the solver works correctly on the refinements.

To check this, we have placed a refined region which covers the spherical overdensity in the above test. The density field on the refinement is chosen to the exactly the same as that on the domain grid, and we fix the boundary condition in the edge cells of the refinement by interpolating from the coarser cells in the domain grid.

The results are shown as red circles in the lower panels of Fig.~\ref{fig:tests}, where we can see that in the domain covered by the refinement they agree with both the solution from numerical integration (black solid curve) and the results on the domain grid (blue squares). We have also done tests using other values of $\delta$ and found similar agreements.

These tests show that the code actually works quite well on both the domain grid and refinements. In the next section, we will show results of cosmological simulations using this code.

\section{Cosmological simulations}

\label{sect:results}

\begin{figure*}
\includegraphics[scale=0.73]{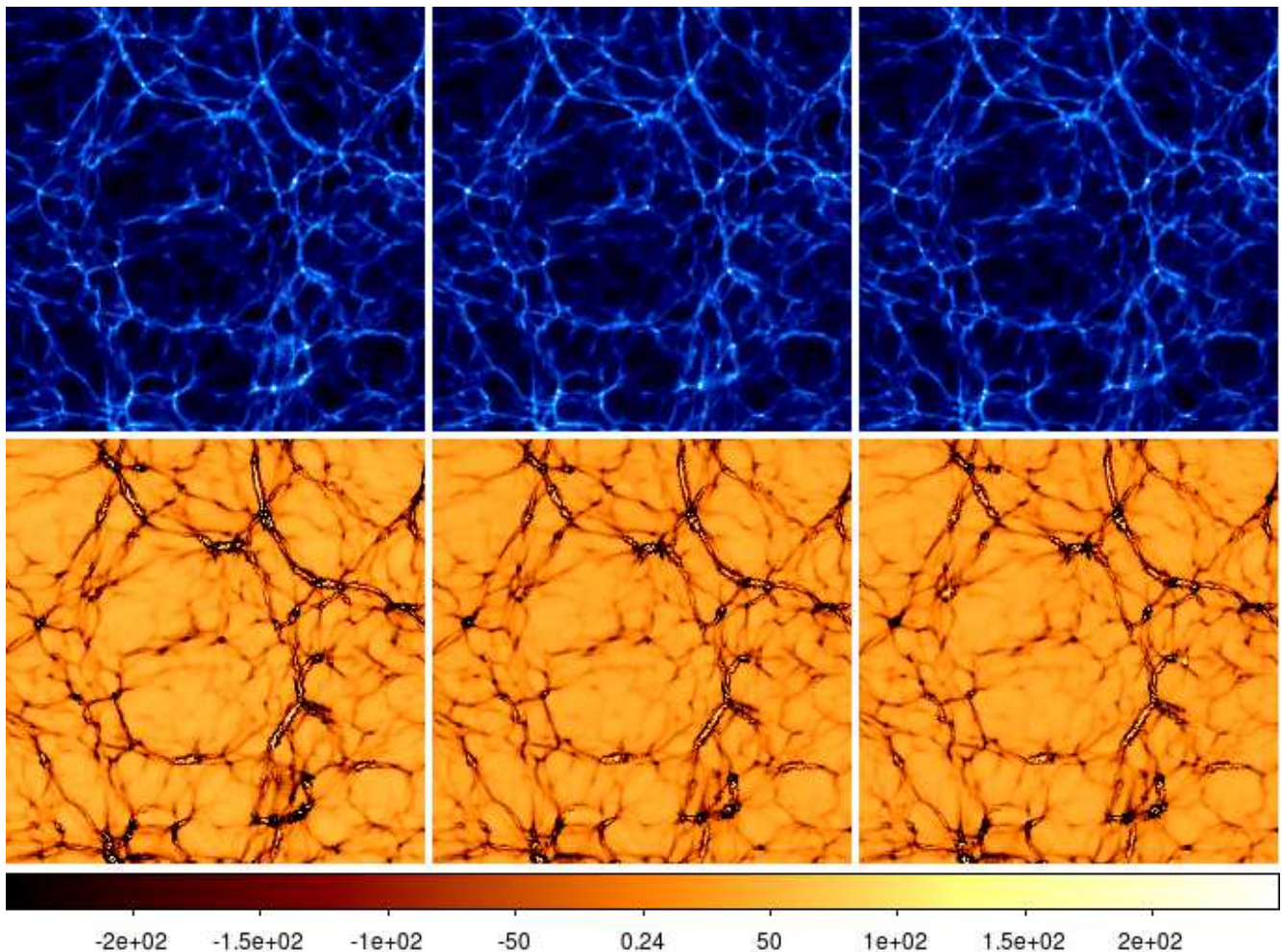}
\caption{(Colour online) Visualisation of the density and velocity divergence fields from our high-resolution simulations with $B=100h^{-1}$Mpc and $256^3$ particles on a domain grid which has 256 cells in each dimension. {\it Upper panels}: the density fields from one realisation of the QCDM (left), linearised DGP (middle) and full DGP (right) simulations, plotted on logarithmic scale; the bright (dark) regions have high (low) matter density. {\it Lower panels:} the velocity divergence fields from the same realisation of our QCDM (left), linearised DGP (middle) and full DGP (right) simulations; the colour bar indicates the values of the velocity divergence. All fields are plotted at $a=1.0$.} \label{fig:visual}
\end{figure*}

To study the cosmological behaviour of the DGP model and the Vainshtein mechanism, we have run a set of 36 $N$-body simulations using the {\sc ecosmog-v} code. In this section we show the results from these simulations.

\subsection{Simulation details}

As we have seen above, the self-accelerating branch of the DGP model shows a different background expansion history from that of $\Lambda$CDM, which itself can contribute to the different matter clustering seen in these two models. Therefore, to see the effects of modified gravitational force and in particular the Vainshtein mechanism more directly, it is better to compare the full DGP simulation with a corresponding Newtonian simulation with the DGP background: this so-called {\it QCDM model} \cite{schmidt2009} is not a rigorous theoretical model and we consider it only for comparisons. Also, we have simulated the {\it linearised DGP model} in which the nonlinear derivative coupling term is set to zero by hand so that the model has exactly the DGP background but particles feel an effective Newton's constant given by
\begin{eqnarray}
G_{\rm eff}(a) &=& \left(1+\frac{1}{3\beta}\right)G.
\end{eqnarray}
Because $\beta<0$ for the self-accelerating branch of the DGP model, the modified gravitational force is actually weaker than in GR.

To make statistic averages, we run six realisations of simulations for every model. The initial conditions are generated using the {\sc mpgrafic} package \cite{mpgrafic} at $a=0.02$, from the same linear matter power spectrum but with different random seeds. Strictly speaking, even at $z_i=49.0$ the matter clustering can be different in QCDM and DGP, but for the self-accelerating branch which we are studying here the difference in the linear matter power spectra is fairly small ($\sim0.2\%$) and so we have neglected it. The fact that we use the same initial conditions for simulations of different models makes sure that the initial density fields have the same phases, and so any difference in the matter power spectra that we find at later times will be a direct consequence of the different dynamics and force laws between the models.

The cosmological parameters we use in the simulations are taken from the best-fitting self-accelerating DGP model using the WMAP 5yr data \cite{fwhhhm2008}. To be more specific, we have
\begin{eqnarray}
&&\left\{h, n_s, \ln\left[10^{10}A_s\right], \Omega_bh^2, \Omega_ch^2, \Omega_{\rm rc}, \sigma_8\right\}\nonumber\\
&=& \{0.66, 0.998, 3.01, 0.0237, 0.0888, 0.138, 0.536\}.\nonumber
\end{eqnarray}
Note that $\Omega_{\rm rc}$ is a derived parameter which makes the universe flat and $\sigma_8$ is its current value for the linearised DGP model (because to compute it uses linear perturbation theory). Some of these parameters, such as $\Omega_{\rm rc}, \Omega_c, \Omega_b$, are used directly in the simulations, while all of them are used in generating the linear matter power spectra at $z_i$.

To better understand the effects of varying the force resolution, we have simulated two different box sizes, respectively $B=100h^{-1}$Mpc and $B=200h^{-1}$Mpc. For both cases we use $256^3$ dark matter particles and the domain grid has 256 cells in each dimension. The cells are refined when the effective number of particles inside them exceeds 9.0 ($N_{\rm pc}=9.0$) and the final refinement level has $2^{14}$ cells in each dimension if they were to cover the whole simulation box. The Gauss-Seidel relaxations stop when $|d^h|<\epsilon=10^{-9}$.

Thanks to the efficient {\sc mpi} parallelisation, the simulations are pretty fast\footnote{Our DGP simulations take roughly 2-3 times CPU time as the corresponding QCDM runs. The simulations are expected to be slower if one increases the resolution, decreases the refinement criterion $N_{\rm pc}$ or decreases the convergence criterion $\epsilon$. Also, in the self-accelerating branch of DGP model considered here, matter clusters less than in QCDM because the fifth force actually weakens gravity; in the normal-branch DGP, on the other hand, gravity is strengthened and matter cluster more strongly, which means there will be more refined grids and therefore the simulation takes longer to finish.}: using 96 CPUs the low-resolution ones finish within 2 hours (wall-clock time) and the high-resolution ones complete in less than 6 hours. The difference is partly because the high-resolution simulations also have more refined time steps, and partly because of the stronger fluctuation of the density field when constructed on a finer grid. Note that although the DGP simulations are in general more difficult to converge than the $f(R)$ simulations (i.e., more tunings of the code are needed), if they do converge they do it much more quickly.

\subsection{Results}

\subsubsection{Visualisation of density and velocity fields}

In Fig.~\ref{fig:visual} we have shown the density and velocity divergence fields in a slice of the simulation box. The velocity divergence field is defined as
\begin{eqnarray}
\theta(\mathbf{x}) &\equiv& \frac{1}{H}\nabla\cdot\mathbf{v}(\mathbf{x}),
\end{eqnarray}
in which $H$ is the Hubble expansion rate and $\mathbf{v}$ is the velocity field of the dark matter particles. To measure the velocity field and its divergence from the particle positions and velocities we have used the Delaunay Tessellation Field Estimator (DTFE) code described in \cite{dtfe}. The Delaunay tessellation has the advantage that the velocity divergence field computed in this way is volume-averaged, rather than mass-averaged. It also avoids the problem of empty cells including no particles, which can arise in direct assignment methods of measuring the velocity field \cite{ps2009}.

Shown in the upper panels of Fig.~\ref{fig:visual} are the logarithmic density field $\log\rho(\mathbf{x})$ for one of the 6 realisations of the QCDM (left), linearised DGP (middle) and full DGP (right) simulations. The colour scale is from $\rho=0.05$ (black) to $\rho=100.0$ (white). The difference in the pattern and strength of clustering is barely visible by eye, but one can still see the stronger clustering in the QCDM model (recall that the in DGP model gravity is weakened compared to QCDM), in particular in the cluster near the lower-right corner. A blink view of these panels also show that in QCDM the clusters are slight closer to each other, while the difference between linearised and full DGP is very small.

Similar patterns can be seen in the lower panels of Fig.~\ref{fig:visual}, in which we show the corresponding velocity divergence fields. The velocity divergence field is positive in low-density regions where matter flows from the central part and the flow becomes faster as it approaches clumps of matter. This trend is reversed near clusters and filaments, where the velocity divergence field becomes negative since matter flows inwards here. Inside clusters and filaments, the velocity divergence takes positive sign again, as noted by \cite{ps2009}. Since gravity is strongest in the QCDM model, the matter flow is faster inside voids and near clusters, making the plot brighter in the low-density regions and the black regions near clusters thicker, though this is barely visible without blinking view (one can however see the difference in the clusters near the lower right corner of each panel).

\subsubsection{Matter and velocity divergence power spectra}

Given that the difference is so visibly small, it is more useful to look at statistical quantities, such as the power spectra, to quantify the matter clustering,
\begin{eqnarray}
P_{\delta\delta}(k) &\equiv& \langle|\delta_{\mathbf{k}}|^2\rangle,
\end{eqnarray}
with
\begin{eqnarray}
\delta_{\mathbf{k}} &\equiv& \left(2\pi\right)^{-3/2}\int\delta(\mathbf{x})\exp(-i\mathbf{k}\cdot\mathbf{x}){\rm d}^3\mathbf{x},
\end{eqnarray}
and
\begin{eqnarray}
P_{\theta\theta}(k) &\equiv& \langle|\theta_{\mathbf{k}}|^2\rangle,
\end{eqnarray}
with
\begin{eqnarray}
\theta_{\mathbf{k}} &\equiv& \left(2\pi\right)^{-3/2}\int\theta(\mathbf{x})\exp(-i\mathbf{k}\cdot\mathbf{x}){\rm d}^3\mathbf{x},
\end{eqnarray}
for the velocity divergence field. In above $\langle\cdot\rangle$ means ensemble average.

\begin{figure*}
\includegraphics[scale=0.475]{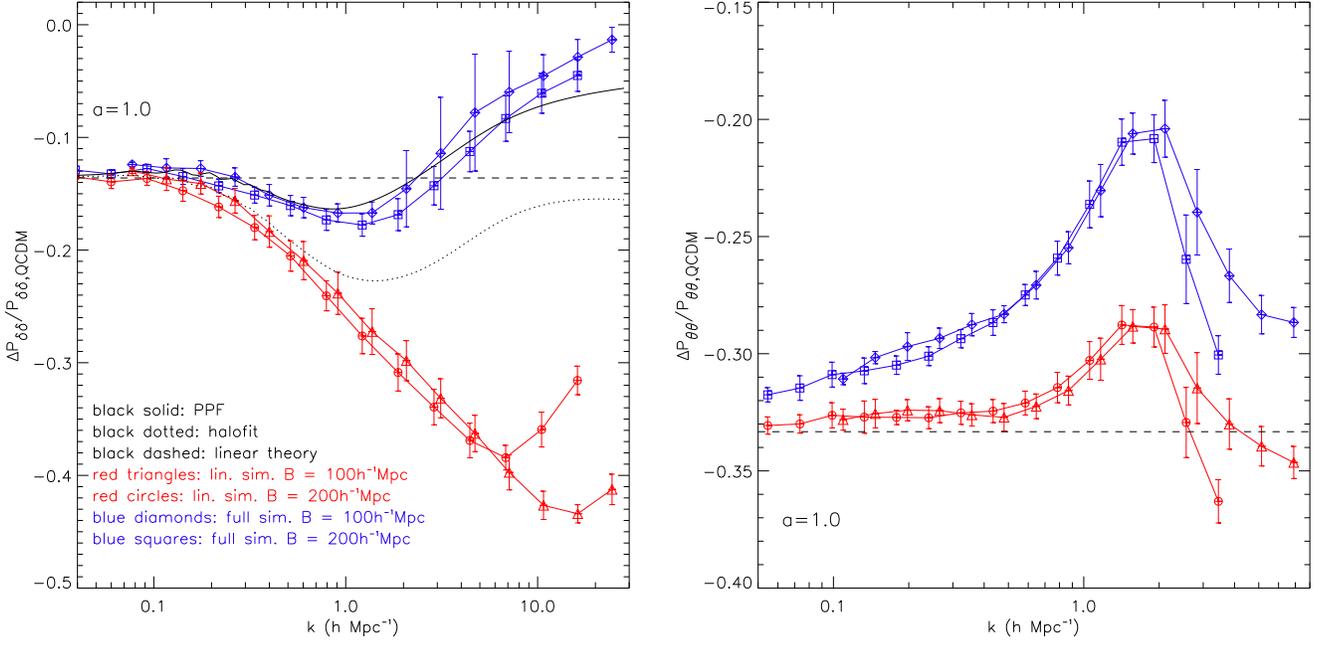}
\caption{(Colour online) The relative differences of the matter (left panel) and velocity divergence (right panel) power spectra of the full (blue symbols) and linearised (red symbols) DGP simulation from that of the QCDM simulation, at $a=1.0$. The results are obtained by averaging 6 realisations for the high- and low-resolution simulations respectively (see the texts in the figure for more details). The velocity divergence is measured from a Delaunay tessellation of the particle distribution. We also show the predictions of linear theory (black dashed curves), Halofit (black dotted curve)  and PPF (black solid curve).} \label{fig:pk_z0}
\end{figure*}

\begin{figure*}
\includegraphics[scale=0.475]{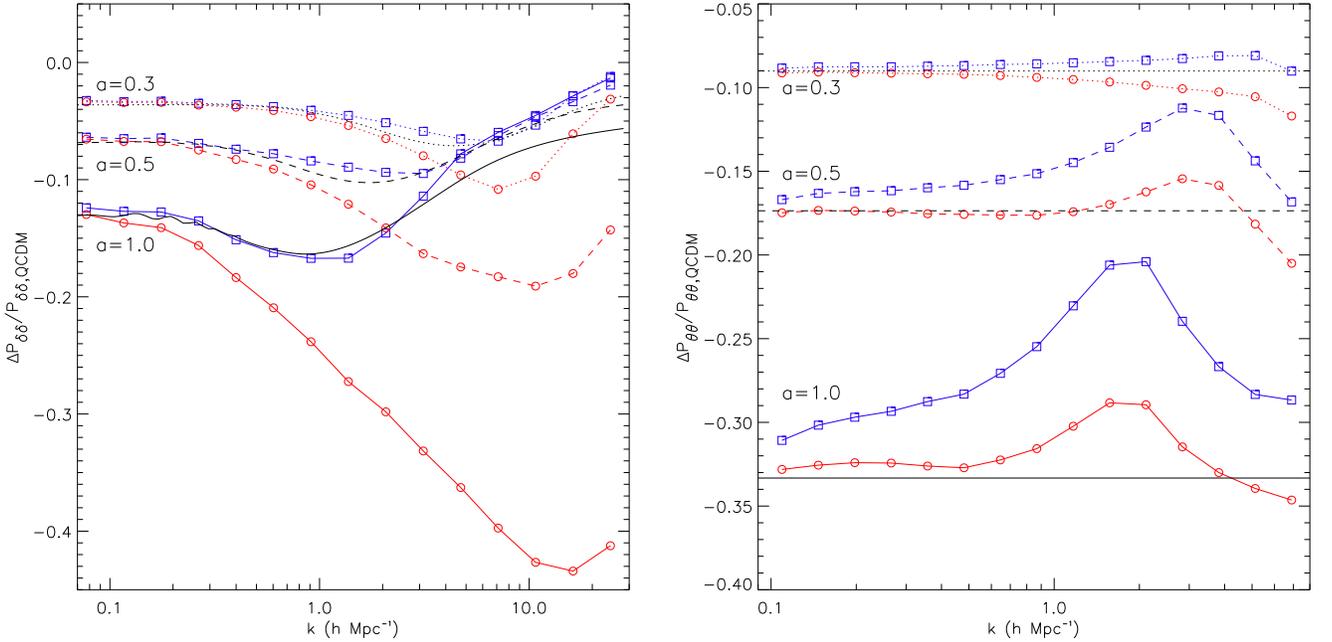}
\caption{(Colour online) Same as Fig.~\ref{fig:pk_z0} but for three different output times -- $a=0.3$ (dotted curves), $a=0.5$ (dashed curves) and $a=1.0$ (solid curves) -- to see the time evolution of the pattern. Blue squares are from the full DGP simulations and red circles from the linearised simulations. For clarify we have only shown the results from the $B=100h^{-1}$Mpc simulations and error bars are not plotted. The smooth black curves in the left (right) panel are predictions of PPF (linear perturbation theory).} \label{fig:pk_evo}
\end{figure*}

In Fig.~\ref{fig:pk_z0} we show the relative difference of $P_{\delta\delta}$ (left panel) and $P_{\theta\theta}$ of the full (blue symbols) and linearised (red symbols) simulations from that of the QCDM simulation. The left panel agrees quite well with previous results such as those in \cite{schmidt2009}, but extends to smaller scales due to the AMR nature of our code. There are several notable features here:
\begin{enumerate}
\item Both the linearised and the full DGP simulations agree with the linear perturbation theory prediction on length scales larger than $k\sim0.1h$~Mpc$^{-1}$. This indicates that, at least for the WMAP5 best-fitting DGP model (the self-accelerating branch), the nonlinearity in Vainshtein mechanism becomes important at $k\sim0.1h$~Mpc$^{-1}$, exactly where the nonlinearity in the underlying density field starts to make linear perturbation theory invalid.
\item On scales smaller than $k\sim0.1h$Mpc$^{-1}$, the Vainshtein effect can strongly suppress the deviation from QCDM compared to the linearised simulation (where it is absent). This agrees with the result of \cite{schmidt2009}.
\item The AMR property of our code enables us to measure the matter power spectra on significantly smaller scales than that of \cite{schmidt2009}. Indeed, the simulations show that the deviation from QCDM in $P_{\delta\delta}$ decays to zero on small scales, in agreement with the halo model prediction of \cite{shl2010}.
\item The low-resolution simulations, with a larger box size, show better agreement with linear perturbation theory on large scales as expected. Box size is an equally, if not more, important issue in our simulations, and should not be chosen to be too small as a compromise to achieve better resolutions. This reflects the importance of AMR from a different angle.
\end{enumerate}

In the framework of GR, fitting formulae such as the Halofit have been developed to provide a mapping between the linear and nonlinear matter power spectra \cite{Halofit}. But due to its simplicity, it is often misused to derive the nonlinear power spectrum in modified gravity models \cite{BBK}. Fig.~\ref{fig:pk_z0} clearly shows that the Halofit overestimates the deviation of DGP from QCDM on small scales. This is hardly surprising because Halofit does not incorporate the effect of the Vainshtein mechanism which helps to recover GR on small scales. The Vainshtein mechanism is effective once the density perturbations become nonlinear \cite{bending}, and brings the matter power spectrum back to the QCDM prediction on small scales. Based on this observation, \cite{PPF} has  suggested a simple way to modify the Halofit to fit the nonlinear power spectrum in modified gravity models, which is called the Parameterised-Post Friedman (PPF) fit. The PPF matter power spectrum is given by
\begin{equation}
P_{\delta \delta}(k, z) = \frac{P_{\delta \delta, \rm non-GR}(k, z) + c_{\rm nl}(z) \Sigma^2(k,z)P_{\delta \delta, \rm QCDM}}
{1+ c_{\rm nl}(z)  \Sigma^2(k,z)},
\end{equation}
where $\Sigma(k,z)$ measures the nonlinearity of the density perturbations
\begin{equation}
\Sigma^2(k, z)= \frac{k^3}{2 \pi^3} P_{\delta \delta, \rm lin},
\end{equation}
$P_{\delta \delta, \rm lin}$ is the linear matter power spectrum for the DGP model and $P_{\delta \delta, \rm non-GR}$ is the power spectrum in the linearised model without the Vainshtein mechanism, which can be obtained by simply applying Halofit to the linear power spectrum for the DGP model. $P_{\delta \delta, \rm QCDM}$ is the power spectrum in the QCDM model. As the nonlinearity becomes large $\Sigma \gg 1$, this formula ensures that the power spectrum in the full DGP model approaches the one in the QCDM model. In Ref.~\cite{PT}, the free parameter $c_{\rm nl}(z)$ in the PPF fit was estimated using the perturbation theory as $c_{\rm nl}(0)=0.3, c_{\rm nl}(1)=0.34$ and $c_{\rm nl}(2.33)=0.36$. Using these predictions for $c_{\rm nl}(z)$, we found that the PPF fit recovers the $N$-body simulation results very well (Figs.~\ref{fig:pk_z0} and \ref{fig:pk_evo}). This demonstrates that the Vainshtein mechanism is working as expected in our full DGP simulations.

The right panel of Fig.~\ref{fig:pk_z0} is the same as the left panel, but for the velocity divergence power spectra. Here, the behaviour is very different. In particular, we see that the Vainshtein mechanism already has effects on sales as large as $k\sim0.05h$Mpc$^{-1}$, similar to what we have seen in our $f(R)$ gravity simulations \cite{lhkzjb2012}. Furthermore, the deviation from QCDM on large scales ($\sim30-35\%$) is much larger than what we see in $P_{\delta\delta}$ ($\sim13\%$ as is shown in the left panel). This means that observationally the velocity field, though more challenging to measure, can be more interesting in the tests of gravity.

\subsubsection{Time evolution of the power spectra}

Finally, Fig.~\ref{fig:pk_evo} shows the time evolutions of the matter (left panel) and velocity divergence (right panel) power spectra. In particular, we show $\Delta P_{\delta\delta}/P_{\delta\delta,{\rm QCDM}}$ and $\Delta P_{\theta\theta}/P_{\theta\theta,{\rm QCDM}}$ at three different epochs, with $a=1.0$ (solid curves), $a=0.5$ (dashed curves) and $a=0.3$ (dotted curves). Results for the full (linearised) simulations are shown in blue squares (red circles) as before.

The left panel of Fig.~\ref{fig:pk_evo} shows that deviations from QCDM are smaller at earlier times, which is because $|\beta|$ is larger and $G_{\rm eff}$ is closer to $G_{N}$ at higher redshifts. The Vainshtein effect is also weaker at the early times because the nonlinear term in the DGP equation is proportional to $|\beta|^{-1}$ and therefore suppressed. However, the general observations that we have for $a=1.0$, notably that the Vainshtein mechanism suppresses the deviations from QCDM on smaller scales and eventually brings things back to GR, still apply here. The evolution pattern can be understood as follows:
\begin{enumerate}
\item Note that the deviation from QCDM essentially freezes on scales smaller than $k\sim0.7h$Mpc$^{-1}$ after $a=0.3$, which shows that those scales become below the Vainshtein radius so that they feel the standard gravity.
\item The very large scales, on the other hand, are beyond the Vainshtein radius until today and their evolution can be approximately described by linear perturbation theory.
\item Going from very large scales to intermediate scales, the nonlinearity of the matter density field first kicks in and makes the full DGP simulation behave similarly to the linearised simulation (i.e., $\Delta P_{\delta\delta}/P_{\delta\delta}$ initially decreases as $k$ increases). Then, the Vainshtein mechanism takes over and brings things back to the small-scale behaviour described above. This explains why $\Delta P_{\delta\delta}/P_{\delta\delta}$ appears to have a valley whose position depends on the cosmic epoch and shifts to larger scales with time.
\end{enumerate}


The velocity divergence power spectrum, on the other hand, shows very different properties from the matter power spectrum:
\begin{enumerate}
\item Even on length scales as large as $k\sim0.1h$Mpc$^{-1}$, $P_{\theta\theta}$ in the full DGP simulations does not agree with linear-theory prediction, though in the linearised DGP simulations it does agree with linear theory on those scales. This implies that the Vainshtein mechanism affects the velocity divergence even on such large scales, and it is in contrast to the matter power spectrum, which agrees with linear theory predictions in both the full and the linearised DGP simulations on those scales. This has an important implication in the measurement of the growth rate using redshift distortions. We need to carefully take into account the Vainshtein effect even on large scales to extract the linear growth rate accurately.
\item On small scales, the Vainshtein mechanism suppresses the deviation from QCDM,a but the velocity divergence power spectrum does not approach to the QCDM result even at $k\sim10h$Mpc$^{-1}$ whereas the matter power spectrum does. This implies that velocities hold a key to distinguish modified gravity models with the Vainshtein mechanism from GR on small scales.
\end{enumerate}

We can also understand the pattern for $\Delta P_{\theta\theta}/P_{\theta\theta}$. Taking the linearised DGP simulation as an example, for which Newton's constant is scaled by $1+1/(3\beta)$ on all scales, so that in linear theory one would expect $\Delta P_{\theta\theta}/P_{\theta\theta}$ to be a constant at least on large scales, as we could see at early times ($a=0.3$). On small scales, the nonlinearity transfers power between different scales, causing the deviation from scale-independence. When clusters start to form, the $\theta$ field changes sign from the outside to the inside of halos, and its magnitude can become smaller. As gravity is stronger in the QCDM model, at a given time it has stronger clustering and more compact clusters, and so the scale at which $\theta$ changes sign is also smaller -- this leads to, on average, an increase in $|\Delta P_{\theta\theta}|/P_{\theta\theta}$ on the typical halo-formation scale and a decrease on slightly larger scale, which gives rise to the peak we see in the plot of $\Delta P_{\theta\theta}/P_{\theta\theta}$. The peak shifts to larger scales because the halo-formation scale increases in time.

\section{Summary and outlook}

\label{sect:summary}

To summarise, in this paper we have reported the development and initial results of a new $N$-body code which is suitable for cosmological simulations in modified gravity theories which restore GR in high-density regions by nonlinear derivative couplings, such as the DGP and Galileon gravity models. This code is based on our {\sc ecosmog} code \cite{ecosmog}, which was originally developed for $f(R)$ gravity simulations and later generalised to simulate generic modified gravity theories which restore GR using nonlinear self-interaction potentials \cite{bdlwz2012,bdlwz2013}. However, the solver for the nonlinear partial differential equation is different in both its algorithm and its implementations, making it essentially a new and independent code.

The code, which we call {\sc ecosmog-v} with the {\sc v} standing for `Vainshtein', has a few distinct features compared with existing codes in the literature which tackle the same problem:
\begin{enumerate}
\item It solves the scalar field on a mesh using nonlinear relaxation, which has good convergence properties \cite{cs2009}. The density field on the mesh is obtained by assigning particles following the triangular-shaped-cloud scheme, and there is no need to pre-smooth it to achieve convergence.
\item The mesh can be adaptively refined in high-density regions to achieve higher resolution and accuracy there, without affecting the overall performance. There is no need to smooth the density field on the refinements either.
\item It is efficiently parallelised using {\sc mpi}, which makes the simulations fast. This is important for modified gravity simulations which generally take much longer than corresponding GR simulations, and makes this code suitable for large and high-resolution cosmological simulations which are essential for us to fully understand the small-scale behaviour of such models, and to better explore the future high-precision observational data on the large-scale structures.
\end{enumerate}

We have made various tests of the code to make sure that it gives accurate solutions to simple situations such as uniform density field, 1D (sine and Gaussian) density fields and spherical overdensites. Because the most important feature of this code is the AMR, we also tested the code on refinements and found that it works correctly. Furthermore, our cosmological simulations predict the same behaviour of the matter power spectrum as that discovered in previous work \cite{schmidt2009,cs2009}.

We have used our code to run a set of 36 simulations for the QCDM, linearised DGP and full DGP models. The AMR nature enables us to probe scales smaller than those reached by previous simulations, and confirmed the halo-model prediction of \cite{shl2010} that the full DGP matter power spectrum should go back to the QCDM result on small enough scales. 

We have also studied the velocity field in DGP gravity, and found that it is more strongly affected by the nonlinearity in both the underlying matter field and the Vainshtein mechanism. Even on large scales where the matter power spectra for the full and linearised DGP simulations agree quite well, we find noticeable difference between the velocity divergence power spectra of the two. This trend starts as early as $z=1$, and by $z=0$ the deviation of $P_{\theta\theta}$ from the corresponding QCDM result can be as large as more than $30\%$, compared with the $\sim13\%$ deviation in $P_{\delta\delta}$. This suggests that large-scale velocity fields can be a good probe of modified gravity theories such as the DGP.

Of course, the DGP model is disfavoured by various observational tests, but the beauty in the idea of Vainshtein mechanism has motivated more general models which recover GR using nonlinear derivative couplings, notably the Galileon models. Recent studies of \cite{blbp2012,blbp2013} have largely advanced our knowledge in the linear perturbation behaviour of these models and found reasonable fits to the CMB and background expansion data, but still leave the open question about the region of validity of linear theory. To fully answer this question is crucial for cosmological tests of the Galileon model, especially in light of the coming high-precision observational data, but this can only be achieved by studying the nonlinear regime of the structure formation numerically. Our code is useful in this context, and could hopefully bring us one important step forward.

\

\begin{acknowledgments}
BL acknowledges supports by the Royal Astronomical Society and Durham University. GBZ is supported by the Dennis Sciama Fellowship at the University of Portsmouth. KK is supported by STFC grant ST/H002774/1, the Leverhulme trust and the European Research Council. The simulations described in this paper were performed on the {\sc cosma} machine at Durham University. We thank Wojtek Hellwing for helps in measuring the velocity divergence power spectra and making the plots.
\end{acknowledgments}

\end{document}